\documentclass[journal]{IEEEtran}
\usepackage[cmex10]{amsmath}
\usepackage{dsfont}
\usepackage{graphicx}
\usepackage{adjustbox}
\usepackage{url, multicol, multirow, color, tikz, enumerate, verbatim, amsfonts,subeqnarray, romannum}
\usepackage{threeparttable}
\usepackage{algorithmic}
\usepackage[ruled,vlined]{algorithm2e}
\usepackage{rotating}
\usepackage{amsmath,bm}
\usepackage{tikz}
\usepackage{tabularx}
\usepackage{stfloats}
\usepackage{hyperref}
\usepackage{booktabs}
\usepackage{enumitem}   
\usetikzlibrary{shapes,arrows,snakes}

\newcommand{\rred}{\textcolor{black}}

\newcommand{\yusub}{\textcolor{black}}

\newcommand{\sred}{\textcolor{black}}

\newcommand{\gred}{\textcolor{black}}
\def\rered{\yu}
\definecolor{darkblue}{rgb}{0.0,0.0,1.0}
\hypersetup{colorlinks,breaklinks,
	linkcolor=darkblue,urlcolor=darkblue,
	anchorcolor=darkblue,citecolor=darkblue}

\def\max{{\text{\textcolor{black}{max}}}}

\newcommand \La{{\Lambda}}

\def\T{{\mathcal T}}
\def\K{{\mathcal K}}
\def\Z{{\mathbb{Z}}}

\def\Cupper{{\overline C}}
\def\Clower{{\underline C}}
\def\Vupper{{\overline V}}
\def\yu{\textcolor{black}}
\def\yutwo{\textcolor{black}}

\definecolor{darkblue}{rgb}{0.0,0.0,1.0}
\hypersetup{colorlinks,breaklinks,
	linkcolor=darkblue,urlcolor=darkblue,
	anchorcolor=darkblue,citecolor=darkblue}

\def\K {{\cal S}_+}

\def\B{{\mathbb{B}}}
\def\R{{\mathbb{R}}}
\def\Z{{\mathbb{Z}}}

\def\V{{V}}

\def\T{{\mathcal T}}
\def\K{{\mathcal K}}

\def\Cupper{{\overline C}}
\def\Clower{{\underline C}}
\def\Vupper{{\overline V}}
\newcommand{\pb}{\overline{P}}

\newcommand{\pl}{\underline{P}}

\definecolor{myblue}{RGB}{3,70,148}

\def\rred{\textcolor{black}}
\def\rered{\textcolor{black}}
\def\guant{\textcolor{black}}

\allowdisplaybreaks
\usepackage{soul}

\newcommand {\Rmnum} [1] {\expandafter\@slowromancap\romannumeral#1\@}

\newtheorem{proposition}{Proposition}

\newtheorem{remark}{Remark}

\usepackage{amsmath}
\newtheorem{theorem}{Theorem}

\newcommand{\guan}{\textcolor{black}}
\newcommand{\gtwo}{\textcolor{black}}

\newcommand{\ignore}[1]{}

\begin{document} \title{An Integral Formulation and Convex Hull Pricing for Unit Commitment}

%\markboth{Submitted to \emph{IEEE Transactions on Power Systems},~June 2019}%~Vol.~6, No.~1, January~2010
%\markboth{Patent Document}%~Vol.~6, No.~1, January~2010
%{Shell \MakeLowercase{\textit{et al.}}: Bare Demo of IEEEtran.cls for Journals}

\author{Yanan Yu, Yongpei~Guan, and Yonghong Chen
%	    ,~\IEEEmembership{Member,~IEEE;}
%	    ,~\IEEEmembership{Student Member,~IEEE}          
%        Yonghong~Chen %~\IEEEmembership{Member,~IEEE;}
      %  Xing~Wang,~\IEEEmembership{Senior Member,~IEEE}

\thanks{Yanan Yu and Yongpei Guan are with the University of Florida, Gainesville, FL 32611. Yonghong Chen \sred{({Consulting} Advisor)} is with the Midcontinent Independent System Operator, Inc. (MISO), Carmel, IN, USA.}
}
\maketitle
\vspace{-2.0cm}
\begin{abstract}
Reducing uplift payments has been a challenging problem for most wholesale markets in US. The main difficulty comes from the unit commitment discrete decision makings. Recently convex hull pricing has {shown promises} to reduce the uplift payments. \yutwo{However, it has been intractable to obtain the optimal convex hull price.} In this \yu{paper}, we describe \yutwo{an} innovative approach to decide the optimal convex hull \yutwo{price} by \yutwo{simply} solving a linear program. 
We also provide an example to illustrate the calculation process. \yutwo{The final computational experiments on a revised IEEE-118 bus system verify the cost} effectiveness by utilizing our proposed approach.      
\end{abstract}


\IEEEpeerreviewmaketitle

\section{Introduction}\label{sec:intro}
For most wholesale markets in US, Independent System Operators (ISOs) collect the bids from the generation and load sides, and then run the unit commitment and economic dispatch problem (UCED) to decide \yu{the uniform market clearing prices (MCPs) or }local marginal prices (LMPs) for transactions. \yu{A traditional \yutwo{approach} is to calculate LMPs as the \yutwo{value of the} dual variables of the system-wide constraints in the linear \yutwo{program relaxation of the UCED problem with the integer variables fixed at their optimal values (called TLMP method hereafter)}~\cite{gribik2007market}}. As indicated in~\cite{gribik2007market,Ruiz2012}, since the UC problem is, in general, a mixed-integer program, in which the convexity is not maintained. Thus, there could be no set of uniform prices that supports a welfare-maximizing solution. For instance, a generation unit could have a ``lost opportunity cost'', which is defined as the gap between a unit's maximum possible profit \yutwo{that could be obtained from self-scheduling based on the given LMPs} and the actual profit obtained by following the ISO's \yu{schedule}. To address this issue, \yutwo{by maintaining} uniform energy prices based on marginal energy costs, ISOs pay the side payments to units, so as to cover their ``lost opportunity costs''. This payment is referred to as ``uplift'' payment. In other words, due to the non-convexity of the UCED problems, there is a positive non-zero gap between the objective value of the primal formulation used by ISOs and the sum of the objectives of the profit maximization models used by each market participant. ISOs need to pay this positive non-zero gap, i.e., \yu{uplift payments}, to the generation side market participants \yu{to motivate their participation into the day-ahead market}. \yutwo{Since uplift payments lead to extra payments from the load side, and non-transparency in the market clearing process, ISOs aim to minimize the uplift payments for their daily operations.} 

To minimize the \yu{uplift payments}, \yu{several pricing schemes have been proposed. In~\cite{ONEILL2005269}, \yutwo{besides the uniform market price calculated in the traditional method}, a set of prices, such as start-up price and capacity price, are derived from the dual variables of the equations, in which the binary variables are fixed at their optimal values in the UCED problem. This pricing provides a Walrasian equilibrium and results in zero profit for all suppliers, while it leads to uplift payments for the generators whose marginal price is less than the market clearing price as indicated in~\cite{gribik2007market}. \yutwo{In~\cite{Zoltowska16}, the author} formulates a direct minimum-uplift model for pricing pool-based auction with network constraints, while the dispatch and commitment-based profits and uplifts are assumed to be decoupled, i.e., no ramping constraints of the generator are \yutwo{considered}.} \yutwo{Besides the above two approaches, }convex hull pricing approach was introduced and \yutwo{also} received significant attention~\cite{gribik2007market,Hogan2003,hua2017convex,schiro2016convex}. This pricing approach aims to minimize the \yu{uplift payments} over all possible uniform prices, \yutwo{and also shows that the uplift payment is essentially equal to the duality gap between the UCED problem and its Lagrangian dual}. \yutwo{Several wholesale markets in US have implemented an approximation of convex hull pricing. For instance, in the }Midcontinent ISO, \yutwo{it is} named extended locational marginal prices \yu{(ELMP)}.

\yutwo{Although convex hull pricing approach is promising in theory as indicated in~\cite{wang2016commitment}, this method requires the optimal Lagrangian multipliers for the mixed-integer UCED problem over the whole operational horizon, which are computationally expensive}. The main difficulties lie in two aspects: 1) discrete decision variables (on and off statuses of each generator) in the formulation and 2) general convex functions of the generation costs. To address these, significant progress has been made in~\cite{hua2017convex}, among others, in which a convex hull description for the UC polytope without considering ramping is introduced and convex envelope is introduced to reformulate the problem as a second-order cone programming. However, the optimal solutions for the uplift payments minimization problem described in \cite{hua2017convex} may not \yutwo{necessarily} be at the extreme points due to ramping constraints. In this paper, we derive an integral formulation, as a complement to the one described in~\cite{guan2018polynomial} for the single-generator UC problem that can provide an integral solution for a general single-generator UC problem considering min-up/-down time, generation capacity, ramping and variant start-up cost restrictions with a general convex cost function~\yu{by solving a linear program}, which could overcome the difficulties. 

In the remaining part of this paper, we describe the integral formulation in Section~\ref{integralformulation}~\yu{and \yutwo{further} provide a complete proof following a dynamic programming framework in Section~\ref{sec:proof}}. In Section~\ref{sec:uplift}, we describe the uplift payments minimization formulation and show how the calculation of the uplift payments minimization problem can be implemented through solving a linear programming problem. Finally, in Section~\ref{sec:exm}, we present an example to explore the insights and \yutwo{report the computational experiment results on a revised IEEE-118 bus system, which verifies the effectiveness of our proposed approach.}

\section{An Integral Formulation for Unit Commitment} \label{integralformulation}
In this section, we introduce an integral formulation for a single-generator UC problem with the generator to be initially on. For a $T$-time period UC problem, the traditional 2-Bin formulation for a single generator can be described as follows:
\begin{subeqnarray} \label{model:Nduc}
	&\hspace{-0.5in}\min  & \sum_{t=1}^{T}f_t(x_t,\rered{u_t}) + \sum_{t=t_0+\ell+1}^{T} \yu{\textcolor{black}{\zeta_t}}  + \sum_{t=t_0}^{T-1} \textcolor{black}{\zeta'_t} \slabel{eqn:Nduc_obj} \\
	&\hspace{-0.5in}\mbox{s.t.}& \hspace{-0.1in}\rered{u_t} = 1, \forall t \in [1, t_0]_{\Z}, \slabel{eqn:Np-initon}\\  
	&& \hspace{-0.5in}\sum_{i = t- L +1}^{t} \rered{v_i} \leq \rered{u_t}, \ \forall t \in [t_0+\ell+L, T]_{\Z}, \slabel{eqn:Np-minup} \\
	&& \hspace{-0.5in}\sum_{i = t- \ell +1}^{t} \rered{v_i} \leq 1 - \rered{u}_{t- \ell}, \ \forall t \in [t_0+\ell, T]_{\Z}, \slabel{eqn:Np-mindn} \\
	&& \hspace{-0.5in}\rered{u_t} - \rered{u_{t-1}} - \rered{v_t} \leq 0, \ \forall t \in [1, T]_{\Z}, \slabel{eqn:Np-udef} \\
	&& \hspace{-0.5in} - x_t + \Clower \rered{u_t} \leq 0, \ \forall t \in [1, T]_{\Z}, \slabel{eqn:Np-lower-bound} \\
	&& \hspace{-0.5in}x_t - \overline{C} \rered{u_t} \leq 0, \ \forall t \in [1, T]_{\Z}, \slabel{eqn:Np-upper-bound} \\
	&& \hspace{-0.5in}x_t - x_{t-1} \leq V \rered{u_{t-1}} + \Vupper (1 - \rered{u_{t-1}}), \ \forall t \in [1, T]_{\Z}, \slabel{eqn:Np-ramp-up} \\
	&& \hspace{-0.5in}x_{t-1} - x_t \leq V \rered{u_t} + \Vupper (1-\rered{u_t}), \ \forall t \in [1, T]_{\Z} \slabel{eqn:Np-ramp-down},\\
	&& \hspace{-0.5in}\textcolor{black}{\zeta'_t} \geq \textcolor{black}{S'(t\yusub{+s_0})}(\yusub{1-u_{t+1}}-\sum_{s=1}^{t}(1-u_s)), \forall t \in [t_0,T-1]_{\Z},\slabel{eqn:Nsd1}\\
	&& \hspace{-0.5in}\textcolor{black}{\zeta'_t} \geq \textcolor{black}{S'(t-k+1)} \bigg(v_k - \sum_{s=k}^{t} (1-u_s) \bigg), \nonumber \\
	&& \hspace{-0.5in} \forall t \in [t_0+\ell+L, T-1]_{\Z}, k \in [t_0+\ell+1, t-L+1]_{\Z},\slabel{eqn:Nsd2} \\
	%&& \textcolor{black}{\zeta'_t} \geq SD(t+s_0)(u_k - \sum_{s=k}^{t}(1-y_s)), \forall t \in [t_0,T-1]_{\Z}, \forall k \in [1,t-L+1]_{\Z},\slabel{eqn:Nsd2}\\
	%&& \textcolor{black}{\zeta_t} \geq SU(t+s_0 -1) \bigg(\rered{v_t} - \sum_{s=1}^{k-1}y_s \bigg),\ \forall t \in [1, T]_{\Z}, \slabel{eqn:Nsu1} \\
	&& \hspace{-0.5in}\textcolor{black}{\zeta_t} \geq \textcolor{black}{S}(t-k-1) \bigg(\rered{v_t} - \sum_{s=k+1}^{t-1}u_s \bigg), \nonumber\\
	&& \hspace{-0.5in} \forall t \in [t_0+\ell +1, T]_{\Z}, k \in [t_0, t-\ell-1]_{\Z},\slabel{eqn:Nsu2} \\
	&& \hspace{-0.5in}\rered{u_t}, \rered{v_t} \in \{0, 1\},\  \textcolor{black}{\zeta_t}, \textcolor{black}{\zeta'_t} \geq 0,  \  u_0=1, \slabel{eqn:Np-nonnegativity}
\end{subeqnarray}
where $L$, $\ell$, $\Cupper$, $\Clower$, $\Vupper$, and $V$ represent the min-up time limit, min-down time limit, the generation upper bound, the generation lower bound, the start-up/shut-down ramp rate, and the ramp-up/-down rate in the stable generation region. The parameters $S(\cdot)$ and $S'(\cdot)$ represent the start-up and shut-down costs determined by the start-up and shut-down times, e.g. $S(t-k+1)$ and $S'(t-k+1)$ represent the start-up and shut-down costs when the generator has been off  and on for $(t-k+1)$ time periods, respectively. \yusub{The decision variables $\rered{u_t}$, $\rered{v_t}$, $x_t$, and $\zeta_t$ represent the generator's on/off status, start-up decision, generation amount, and start-up cost at time $t$, respectively. The decision variable $\zeta'_t$ represents the shut-down cost when the generator shuts down at time $t+1$.} The objective is to minimize the net total cost where $f(\cdot)$ denotes the fuel cost minus the revenue as a function of its electricity generation amount and electricity price. In this study, we assume $f(\cdot)$ is \rered{piecewise linear} with respect to $x$ and the generator has been initially on for $s_0 $ time periods before time period $1$. Then, the generator cannot shut down until time $t_0+1$, \guan{with $t_0 = [L-s_0]^+$,} due to min-up constraints (as indicated in constraint~\eqref{eqn:Np-initon}). \gtwo{For notation convenience, we use $[a,b]_\Z$ to represent the set of integers $\{a,a+1,\dots,b\}$ \yusub{and the set $[a,b]_\Z$ is empty if $a>b$}.}

This traditional formulation could not provide an integral solution by solving it as a linear problem (i.e., relaxing the integrality constraints for the binary variables). In this \guan{section}, we illustrate an alternative formulation which could provide an integral solution for the binary variables by solving the problem as a linear program. We first introduce the formulation as follows and then provide the proof in the remaining part of this section.

For the alternative formulation, we keep track \guan{of} the consecutive ``on'' and ``off'' periods of each generator. We let binary decision variable \yu{$w_{t}, t = 1,\dots,T-1$,} represent whether the generator shuts down at time $t+1$ for the first time ($w_{t}=1$) or not ($w_{t}=0$), \yu{$w_T$ shows whether the generator keeps on for the whole time period, }binary decision variable {\yu{$y_{tk}, k \leq T-1$,}} represent whether the generator starts up at $t$ and shuts down at $k+1$ ($y_{tk}=1$) or not ($y_{tk}=0$), \yu{$y_{tT}$ represents whether the generator starts up at $t$ and keeps on till the end of the horizon}, binary decision variable {$z_{tk}$} represent whether the generator shuts down at time $t+1$ and starts up again at time $k$ ($z_{tk}=1$) or not ($z_{tk}=0$), and binary decision variable $\theta_t$ represent whether the generator shuts down at time $t+1$ and stays offline to the end ($\theta_t =1$) or not ($\theta_t =0$). We also let $q_{tk}^{s}$ be the generation amount at time $s$ corresponding to the ``on'' interval $[t, k]_{\Z}$. The corresponding \yusub{net cost} ${f_{tk}^s(q_{tk}^{s})}$ is usually denoted as a convex function ${\bar{f}_{tk}^s(q_{tk}^{s})}=\yusub{\alpha}_s(q_{tk}^{s})^2+\yusub{\beta}_sq_{tk}^{s}+c_s$ if only generation cost is involved or a convex function ${f_{tk}^s(q_{tk}^{s})}=\yusub{\alpha}_s(q_{tk}^{s})^2+(\yusub{\beta}_s-\textcolor{black}{p}_s)q_{tk}^{s}+c_s$ if the electricity price is also involved, say e.g., $\textcolor{black}{p}_s$ per unit power. Note here that ${\bar{f}_{tk}^s(q_{tk}^{s})}$ is a special case of ${{f}_{tk}^s(q_{tk}^{s})}$ with $\textcolor{black}{p}_s=0$. \yusub{As indicated above for $f(\cdot)$, $f_{tk}^s(q_{tk}^{s})$ is also approximated by a piecewise linear function.} The corresponding formulation, \gtwo{named EUC formulation,} can be shown as follows:
\begin{subeqnarray} \label{model:LP_D2}
	&\min  &  \sum_{t=t_0}^{T-1} \textcolor{black}{S'}(t\yusub{+s_0})\textcolor{black}{w_t} +\hspace{-0.1in} \sum_{t=t_0+\ell+1}^{T-L}\sum_{k =t+L-1}^{T-1}\hspace{-0.1in} \textcolor{black}{S'}(k-t+1) \textcolor{black}{y_{tk}} \nonumber\\
	&& + \sum_{kt \in \K\T} \textcolor{black}{S}(t-k-1) \textcolor{black}{z_{kt}} + \sum_{tk \in \T \K }  \sum_{s=t}^{k} \phi_{tk}^s  \slabel{eqn:LP_D2obj}\\
	%&& \ \sum_{k=t_0}^{T-\ell-1}\sum_{t=k+\ell+1}^{T} \textcolor{black}{S} \textcolor{black}{z_{kt}} + \sum_{tk \in \T \K }  \sum_{s=t}^{k} \phi_{tk}^s  \slabel{eqn:LP_D1obj}\\
	&\mbox{s.t.}  & \sum_{t=t_0}^{T} \textcolor{black}{w_t} = 1, \slabel{eqn:LP_D21}\\
	&& \hspace{-0.5in}-\textcolor{black}{w_t} + \hspace{-0.1in} \sum_{k=t+\ell + 1}^{T}\hspace{-0.1in} \textcolor{black}{z_{tk}} - \hspace{-0.2in}\sum_{k=t_0+\ell+1}^{t-L+1}\textcolor{black}{y_{kt}} +\theta_t = 0, \forall t \in [t_0, T-1]_{\Z},\slabel{eqn:LP_D22}\\
	&& \hspace{-0.5in}	\sum_{k=\min\{t+L-1,T\}}^{T}\hspace{-0.1in}\textcolor{black}{y_{tk}} -\hspace{-0.1in} \sum_{k=t_0}^{t-\ell -1} \textcolor{black}{z_{kt}}= 0, \forall t \in [t_0+\ell+1, T]_{\Z},\slabel{eqn:LP_D23}\\
	&& \hspace{-0.5in}\pl\textcolor{black}{w_k} \leq q_{tk}^s \leq \pb\textcolor{black}{w_k}, \ \forall s \in [t,k]_\Z, \forall tk \in \T\K^1,  \slabel{eqn:LP_D251}\\
	&& \hspace{-0.5in}\pl\textcolor{black}{y_{tk}} \leq q_{tk}^s \leq \pb\textcolor{black}{y_{tk}}, \ \forall s \in [t,k]_{\Z}, \forall tk \in \T\K^2, \slabel{eqn:LP_D252}\\
	&& \hspace{-0.5in}q_{tk}^t \leq \overline{V}\textcolor{black}{y_{tk}}, \ \forall tk \in \T\K^2, \slabel{eqn:LP_D26}\\
	&& \hspace{-0.5in}q_{tk}^k \leq \overline{V}\textcolor{black}{w_k}, \ \forall tk \in \T\K^1, k\leq T-1 \slabel{eqn:LP_D261}\\
	&& \hspace{-0.5in}q_{tk}^k \leq \overline{V}\textcolor{black}{y_{tk}}, \ \forall tk \in \T\K^2, k\leq T-1 \slabel{eqn:LP_D262}\\
	&& \hspace{-0.5in}q_{tk}^{s-1} - q_{tk}^s \leq\V \textcolor{black}{w_k}, \ q_{tk}^{s} - q_{tk}^{s-1} \leq\V \textcolor{black}{w_k}, \nonumber\\
	&& \hspace{-0.5in} \forall s \in [t+1,k]_{\Z}, \forall tk \in \T\K^1, \slabel{eqn:LP_D27}\\
	&& \hspace{-0.5in}q_{tk}^{s-1} - q_{tk}^s \leq\V \textcolor{black}{y_{tk}}, \  q_{tk}^{s} - q_{tk}^{s-1} \leq\V \textcolor{black}{y_{tk}}, \nonumber\\
	&& \hspace{-0.5in} \forall s \in [t+1,k]_{\Z}, \forall tk \in \T\K^2, \slabel{eqn:LP_D28}\\
	&& \hspace{-0.5in}\phi_{tk}^s \hspace{-0.05in}-\hspace{-0.05in} \yusub{a_j^s q_{tk}^{s}}\hspace{-0.05in} \geq \hspace{-0.05in}\yusub{b_j^s} \textcolor{black}{w_k}, \forall s \in [t,k]_{\Z}, j \in  [1,N]_{\Z}, \forall tk \in \T\K^1, \slabel{eqn:LP_D291}\\
	&& \hspace{-0.5in}\phi_{tk}^s\hspace{-0.05in} -\hspace{-0.05in} \yusub{a_j^s q_{tk}^{s}}\hspace{-0.05in} \geq\hspace{-0.05in} \yusub{b_j^s} \textcolor{black}{y_{tk}}, \forall s \in [t,k]_{\Z}, j \in  [1,N]_{\Z}, \forall tk \in \T\K^2, \slabel{eqn:LP_D292}\\
	&& \hspace{-0.5in}\textcolor{black}{w}, \textcolor{black}{z},y \geq 0, \theta_t \geq 0, \forall t \in [t_0,T-\ell-1]_\Z,\slabel{eqn:LP_D293}
	%&& \hspace{-0.5in}\forall k \in [t+L-1, T-1]_{\Z}, \forall t \in [t_0+\ell+1, T-L]_{\Z}. \slabel{eqn:LP_D292}
\end{subeqnarray}
where $ \T\K $ \guan{= $\T\K ^1 \cup \T\K^2$ with} $\T\K^1$ representing the set of all possible combinations of $ t = 1$ and each $k \in [t_0+1,T]_\Z$ to construct a time interval $[t,k]_{\Z}$, and $\T\K^2$ representing the set of all possible combinations of each $t \in [t_0+\ell+1, T]_{\Z}$ and each $k \in [\min\{t+L-1,T\}, T]_{\Z}$ to construct a time interval $[t,k]_{\Z}$. $ \K\T $ represents the set of all possible combinations of each $k \in [t_0, T-\ell -1]_{\Z}$ and each $t \in [k+\ell+1, T]_{\Z}$ to construct a time interval $[k,t]_{\Z}$. Constraints~\eqref{eqn:LP_D21}-\eqref{eqn:LP_D23} keep track of the ``on'' and ``off'' statuses of the unit. Constraints~\eqref{eqn:LP_D251}-\eqref{eqn:LP_D252} represent the generation upper and lower bounds. Constraints~\eqref{eqn:LP_D26}-\eqref{eqn:LP_D262} represent the start-up and shut-down ramping restrictions. Constraints~\eqref{eqn:LP_D27}-\eqref{eqn:LP_D28} represent the ramp-up and ramp-down restrictions. Constraints~\eqref{eqn:LP_D291}-\eqref{eqn:LP_D292} represent the \guan{constraint form for the }piecewise linear convex function ${f_{tk}^s(q_{tk}^{s})}$. 

%\yusub{Note here, the sum items, such as $\sum_{k=t\hspace{-0.01in}+\ell \hspace{-0.01in}+ 1}^{T} \textcolor{black}{z_{tk}}$, will be equal to 0, if the lower index $k\hspace{-0.02in}=\hspace{-0.02in}t\hspace{-0.02in}+\hspace{-0.02in}\ell\hspace{-0.02in}+\hspace{-0.02in}1$ is greater than the upper index $T$.}

\textcolor{black}{
For notation brevity, in the next section, we define the feasible region describing constraints~\eqref{eqn:LP_D21} to~\eqref{eqn:LP_D293} as set $\mathcal{X}_I$. That is,
$\mathcal{X}_I=\{(\gred{w}, \phi, \theta, \gred{y}, \gred{z}, \gred{q}): \mbox{Constraints}~\eqref{eqn:LP_D21}-\eqref{eqn:LP_D293}\}$. In general, network flow formulations 
with side constraints and convex cost functions cannot guarantee an integral solution. However, due to the problem structure, \guan{this above alternative} formulation can provide an integral solution based on the strong duality proof. \guan{In the following section, we provide the corresponding proof.}
%as long as the cost function corresponding to each on interval $[t, k]$ is a convex function of the generation amounts.
}

\section{\guan{The dynamic programming result and duality relationship}} \label{sec:proof}
\guan{For the main part, we show that the EUC formulation is the dual formulation of the primal dynamic programming formulation. Now we describe the dynamic programming formulation first. \yu{W}ithout loss of generality, we assume }the generator has been \guan{initially} online for $s_0$ time periods. \guan{Accordingly, the earliest \guant{shut-down} time will be $t_0+1$ with $t_0 = [L-s_0]^+$. To make the dynamic programming more efficient, we define two value functions, $V_\downarrow(t)$ and $V_\uparrow(t)$ for each time periods. More specially, }we let $V_\downarrow (t)$ represent the \yusub{net cost (generation cost minus revenue)} from time $t$ to the end when the generator shuts down at time $t +1$ (i.e., $t$ is the last ``on'' period for the current ``on'' interval),%, as shown in Figure \ref{fig:1st_value}, 
and $ V_\uparrow (t) $ represent the \yusub{net} cost from time $ t $ to the end when the generator starts up at time $ t $. \guan{Since the generator will be on for consecutive time periods followed by consecutive ``off" time periods, we let $V_\downarrow(t)$ and $V_\uparrow(t)$ keep track of the starting points for the ``OFF" and ``ON" intervals. }% as shown in Figure \ref{fig:2nd_value}. 
\rered{Accordingly, we let $C(t, k)$ represent the \yusub{net cost} \guan{in the interval $[t,k]_\Z$ for the generator} when the generator starts up at time $t$ and shuts down at time $k+1$, and $C(t,T)$ represents the \yusub{net cost} when the generator starts up at time $t$ and keeps online until the end \guan{of the horizon}}. Thus, we have the following dynamic programming equations:
\begin{subeqnarray} \label{new_dp1}
	%&& V_\uparrow (t) = \min_{  \substack{ k \in [\min \{t+L-1, \\  T-1\}, T-1]_{\Z} }  } \bigg\{ SD(k-t+1) + C(t,k) +  V_\downarrow (k), C(t,T) +  V_\downarrow (T) \bigg\}, \ \forall t \in [1, T]_{\Z},  \slabel{eqn:Ndp1} \\
	& V_\downarrow (t) = & \hspace{-0.05in}\min_{ k \in [t+\ell+1,T]_{\Z} } \hspace{-0.03in} \{ \textcolor{black}{S}(k-t-1)+  V_\uparrow (k),  0 \}, \nonumber\\
	&& \hspace{1in} \forall t \in [t_0, T-\ell -1]_{\Z},  \slabel{eqn:Ndp1} \\
	&V_\downarrow (t) = &0, \ \forall t \in [T-\ell, T]_{\Z}, \slabel{eqn:Ndp1p}\\
	&V_\uparrow (t) = & \min_{  \substack{ k \in [t+L-1, T-1]_{\Z} }  } \{ \textcolor{black}{S'}(k-t+1) + C(t,k) +  V_\downarrow (k),\nonumber\\
	&& C(t,T) \}, \quad  \forall t \in [t_0+\ell+1, T-L]_{\Z},  \slabel{eqn:Ndp2} \\
	&V_\uparrow (t) = &C(t,T), \hspace{0.45in}\forall t \in [T-L+1,T]_{\Z}\gtwo{,} \slabel{eqn:Ndp3}
\end{subeqnarray}
%where $ C(t,k) $ represents the optimal generation cost (i.e., the objective value of economic dispatch problem) if the {generator} starts up at time $ t $ and shuts down at time $ k + 1$ (i.e., online at $k$).
\guan{where }equations \eqref{eqn:Ndp1} indicate that when the {generator} shuts down at time $t+1$, it can either start up again at time $k$ with $ k-t-1 \geq \ell $ and $k \leq T$ or keep offline throughout all the remaining time periods. Equations \eqref{eqn:Ndp2} indicate that when the {generator} starts up at time $t$, it can either keep online until time $k$ with $k-t+1 \geq L$ and $k \leq T-1$ or keep online throughout all the remaining time periods. Following the shut-down (resp. start-up) profile, our shut-down (resp. start-up) function can capture the length of online (resp. offline) time before shutting down (resp. starting up). Equations \eqref{eqn:Ndp1p} describe that the {generator} cannot start up again after {it shuts} down at $t+1$ with $T-\ell \leq t \leq T$ due to the \guan{min}-down time limit and equations~\eqref{eqn:Ndp3} describe that the generator must keep online throughout all the remaining time periods if it starts up at time $t$ with $T-L+1 \leq t \leq T$ due to the \guan{min}-up time limit.

As we consider the deterministic UC problem from times $1$ to $T$ and assume the {generator} has been online for $ s_0 $ time periods, our goal is to find out the value of the following function:
\begin{eqnarray} \label{new_dp2}
z = V_\uparrow (\guan{0}) := \hspace{-0.2in} \min_{t \in [t_0,T-1]_{\Z}} \hspace{-0.05in}\Big\{\textcolor{black}{S'}(\yusub{t+s_0}) + C(1,t)+ V_\downarrow (t), C(1,T)\Big\}.
\end{eqnarray}

The corresponding optimal solution can be obtained by tracing the candidates for each optimal value function \guan{backwards from $V_\uparrow(T)$}.%, and this takes $ \Oe (T^3) $ time in total.
%In order to obtain the optimal objective value and corresponding optimal solution, we calculate $V_\uparrow (t)$ and $V_\downarrow (t)$ for all $t$ and record the optimal candidates for them. To calculate the value of each optimal value function in Bellman equations \eqref{eqn:Ndp1} -- \eqref{eqn:Ndp3} {corresponding to each} $t \leq T$, we search among the candidate solutions for each {possible} $k \leq T$, which takes $ \Oe(T) $ time. Thus, the total time to calculate $V_\uparrow (-s_0) $ is $ \Oe (T^2) $ for the aforementioned ``part II''. The corresponding optimal solution can be obtained by tracing the optimal candidates for each optimal value function starting from $V_\uparrow (-s_0)$, and this takes $ \Oe (T) $ time in total. %In summary, our backward induction dynamic programming algorithm for the deterministic UC problem takes $ \Oe (T^2) $ time for ``part II'' (i.e., if all $ C(t,k) $ are presolved), meaning that our algorithm refines the algorithm in \cite{frangioni2006solving}. More importantly, our algorithm is {beneficial} to derive a better reformulation in the following section.

%\subsection{An Extended Formulation for Deterministic UC with \rred{Piecewise Linear Cost Function}}
\guan{In this paper, instead of using dynamic programming to solve the problem efficiently, we use it to derive }an extended formulation for the deterministic UC problem.
By incorporating the dynamic {programming} equations (i.e., \eqref{eqn:Ndp1} - \eqref{eqn:Ndp3} and  \eqref{new_dp2}) as constraints, we obtain the following equivalent linear program:
\begin{subeqnarray} \label{model:LP}
	&\max  &\Phi \slabel{eqn:LP_obj}\\
	\rred{(\textcolor{black}{w_t})} 	&\hspace{-0.3in}\mbox{s.t.} & \hspace{-0.3in}\guan{\Phi} \leq \textcolor{black}{S'}(t\yusub{+s_0}) \hspace{-0.05in} + C(1,t)+ V_\downarrow (t), \hspace{-0.05in} \forall t \in [t_0,T\hspace{-0.05in}-\hspace{-0.05in}1]_{\Z}, \slabel{eqn:LP1}\\
	(\textcolor{black}{w_t}) & & \hspace{-0.3in} \guan{\Phi} \leq C(1,T), \slabel{eqn:LP2}\\
	\rred{(\textcolor{black}{z_{kt}})} 	&&\hspace{-0.3in} V_\downarrow (k) \leq \textcolor{black}{S}(t-k-1) +  V_\uparrow (t), \nonumber\\
	&& \ \forall t \in [k+\ell +1,T]_{\Z}, \forall k \in [t_0, T\hspace{-0.05in}-\ell -\hspace{-0.05in}1]_{\Z}, \slabel{eqn:LP3}\\
	(\theta_t) && \hspace{-0.3in}V_\downarrow (t) \leq 0, \ \forall t \in [t_0, T-\ell -1]_{\Z},\slabel{eqn:LP3P}\\
	\rred{(\theta_{t})}	&&\hspace{-0.3in} V_\downarrow (t) = 0, \ \forall t \in [T-\ell, \guan{T}]_{\Z},\slabel{eqn:LP3pp}\\
	\rred{(\textcolor{black}{y_{tk}})} 	&  &\hspace{-0.3in} V_\uparrow (t) \leq  \textcolor{black}{S'}(k-t+1) +  C(t,k) +  V_\downarrow (k), \nonumber \\
	&&  \hspace{-0.5in} \forall k \in [t+L-1, T-1]_{\Z}, \forall t \in [t_0+\ell+1, T-L]_{\Z},  \slabel{eqn:LP2}\\
	\rred{(\textcolor{black}{y}_{tT})} 	&&\hspace{-0.3in} V_\uparrow (t) \leq  C(t,T) , \ \forall t \in [t_0+\ell+1, T-L]_{\Z}, \slabel{eqn:LP22P}\\	
	\rred{(\textcolor{black}{y}_{tT})} 	&& \hspace{-0.3in}V_\uparrow (t) =  C(t,T) , \ \forall t \in [T-L+1, T]_{\Z}\gtwo{.} \slabel{eqn:LP4}
\end{subeqnarray}

Note here that the optimal value functions in the dynamic program \guan{framework }become decision variables in the above formulation. {To} obtain the value $V_\uparrow (0) $ under the dynamic programming framework, it is equivalent to maximizing variable $ z $ in the linear program above.

Since the above linear program cannot be solved directly as $ C(t,k) $ (the objective value of {the} economic dispatch problem) is unknown \guan{for each pair ($t,k$)}, we first show how to obtain the value of $ C(t,k) $ by discussing two cases: (\romannumeral 1) $ k \leq T-1 $ and (\romannumeral 2) $ k = T $, respectively.

When $ k \leq T-1 $, \guan{if $t>1$, }we have the following formulation to calculate $C(t,k)$ with $(t,k)$ given:
\begin{subeqnarray} \label{model:ED}
\hspace{-0.2in}	C(t,k) = & \hspace{-0.1in}\min & \sum_{s=t}^{k}  \phi_s  \slabel{eqn:ED_obj} \\
\hspace{-0.2in}	\rred{(\lambda_s^-)} \hspace{0.2in}	& \mbox{s.t.}\hspace{0.2in}& \hspace{-0.3in}-x_s \leq -\underline{C}, \ \forall s \in [t,k]_{\Z}, \slabel{eqn:EDlow}\\
\hspace{-0.2in}	\rred{(\lambda_s^+)} \hspace{0.2in}	&&\hspace{-0.3in} x_s \leq \overline{C}, \ \forall s \in [t,k]_{\Z}, \slabel{eqn:EDupp}\\
\hspace{-0.2in}	\rred{(\mu_t)} \hspace{0.2in}	&& \hspace{-0.3in}x_t \leq \overline{V}, \slabel{eqn:EDram1}\\
\hspace{-0.2in}	\rred{(\mu_k)} \hspace{0.2in}	&&\hspace{-0.3in} x_k \leq \overline{V}, \slabel{eqn:EDram2}\\
\hspace{-0.2in}	\rred{(\sigma_s^-)} \hspace{0.2in}	&&\hspace{-0.3in} x_s - x_{s-1} \leq V, \ \forall s \in [t+1,k]_{\Z}, \slabel{eqn:EDramup}\\
\hspace{-0.2in}	\rred{(\sigma_s^+)} \hspace{0.2in}	&& \hspace{-0.3in}x_{s-1} - x_s \leq V, \ \forall s \in [t+1,k]_{\Z}, \slabel{eqn:EDramdo}\\
\hspace{-0.2in}	\rred{(\delta_{sj})} \hspace{0.2in}	&& \hspace{-0.3in}\phi_s \geq a_j^s x_s + \rered{b_j^s}, \ \forall s \in [t,k]_{\Z}, j\in  [1,N]_{\Z}. \slabel{eqn:EDlin}
\end{subeqnarray}

\rered{\guan{If} $t=1$, we have the corresponding formulation by removing constraint~\eqref{eqn:EDram1} as the generator is \guan{initially} on from the initial state}. {When} $ k = T $, we {have the corresponding formulation by removing} constraint \eqref{eqn:EDram2} as the {generator} is not required to shut down at time $ T+1 $ if it stays online until time $ T $. Note here {that} we \guan{have }assumed the net cost function to be piecewise linear. \guan{It could be} indicated by constraints \eqref{eqn:EDlin} with $N$ pieces. {More specifically}, here we use continuous variable $\phi_s$ to represent the \yusub{net cost} at time $s$, while $a_j^s$ and $\rered{b_j^s}$ are the slope and intercept of the $j$th piece of the \yusub{net cost} function at time $s$, respectively.

\guan{Now, we can} incorporate the economic dispatch constraints (e.g., \eqref{eqn:EDlow} - \eqref{eqn:EDlin}) into our proposed linear program \eqref{model:LP}, we take the dual of the economic dispatch model~\eqref{model:ED} and embed its dual formulation into model \eqref{model:LP}. For instance, for  $ k \leq T-1 $, we have the dual formulation as follows.
\begin{subeqnarray} \label{model:EDd}
	C(t,k) = & \hspace{-0.1in}\max &\hspace{-0.1in} \sum_{s=t}^{k} \bigg(\lambda_s^+ \overline{C} - \lambda^-_s \underline{C} \bigg) + \overline{V} (\mu_t+\mu_k) \nonumber\\
	&& + \sum_{s=t+1}^{k} V \bigg( \sigma_s^+ + \sigma_s^- \bigg) + \sum_{s=t}^{k}\sum_{j=1}^{N} \rered{b_j^s} \delta_{sj} \slabel{eqn:EDd_obj} \\
	\rred{(q_{tk}^t)} \hspace{0.1in}	&\hspace{-0.3in}\mbox{s.t.} & \hspace{-0.3in}\lambda_t^+ - \lambda_t^- + \mu_t - \sigma_{t+1}^- + \sigma_{t+1}^+ - \sum_{j=1}^{N} a_j^t \delta_{tj} = 0, \slabel{eqn:EDd1}\\
	\rred{(q_{tk}^k)} \hspace{0.1in}	&& \hspace{-0.3in}	\lambda_k^+ - \lambda_k^-  +\mu_k + \sigma_{k}^- - \sigma_{k}^+ - \sum_{j=1}^{N} a_j^k \delta_{kj} = 0, \slabel{eqn:EDd2}\\
	\rred{(q_{tk}^s)} \hspace{0.1in}	&& \hspace{-0.3in}	\lambda_s^+ - \lambda_s^-  + \sigma_{s}^- - \sigma_{s+1}^-  - \sigma_{s}^+ + \sigma_{s+1}^+ \nonumber \\
	&& - \sum_{j=1}^{N} a_j^s \delta_{sj} = 0,  \forall s \in [t+1,k-1]_{\Z}, \slabel{eqn:EDd3}\\
	\rred{(\phi_{tk}^s)} \hspace{0.1in}	&&\hspace{-0.3in} \sum_{j=1}^{N} \delta_{sj} = 1, \ \forall s \in [t,k]_{\Z}, \slabel{eqn:EDd4}\\
	&& \hspace{-0.3in}\lambda_s^\pm \leq 0, \ \forall s \in [t,k]_{\Z}, \nonumber\\
	&& \hspace{-0.3in} \mu_t \leq 0, \ \mu_k \leq 0, \ \sigma_{s}^\pm \leq 0, \ \forall s \in [t+1,k]_{\Z}, \nonumber\\
	&& \hspace{-0.3in}\delta_{sj}\geq 0, \ \forall j \in [1,N]_{\Z}, s \in [t,k]_{\Z}, \slabel{eqn:EDd5}
\end{subeqnarray}
where $ \lambda_s^- $ and $ \lambda_s^+ $ are dual variables corresponding to constraints \eqref{eqn:EDlow} and \eqref{eqn:EDupp}, respectively, $ \mu_t $ and $ \mu_k $ are the dual variables corresponding to constraint \eqref{eqn:EDram1} and \eqref{eqn:EDram2}, respectively, $\sigma_{s}^- $ and $ \sigma_{s}^+ $ are dual variables corresponding to constraints \eqref{eqn:EDramup} and \eqref{eqn:EDramdo}, respectively, and $ \delta_{sj} $ are dual variables corresponding to constraints \eqref{eqn:EDlin}. \rred{Note that all of these dual variables are correspondingly labeled in the brackets of the left hand side of \eqref{model:ED} for an easy reference.} For $k = T$, we obtain the corresponding dual formulation by removing the dual variable $ \mu_k $ from model \eqref{model:EDd}. \rred{Thus, in the following part of this section, we refer to \eqref{model:EDd} as the dual formulation for all possible $k$, where $ \mu_k $ will be removed from \eqref{model:EDd} when $k=T$.}
Now we obtain an \guan{alternative} linear program, as shown in the following, by plugging the dual formulation of {the} economic dispatch problem and {redefining} $C(t,k)$ to be a decision variable in the following model.
%plug in the dual economic dispatch formulation to our linear program by adding the constraints \eqref{eqn:EDd1} -- \eqref{eqn:EDd5} to \eqref{model:LP} in the following way:
\begin{subeqnarray} \label{model:LP_ED}
	& \max  &  \guan{\Phi} \slabel{eqn:LP_EDobj}\\
	& \mbox{s.t.}  &  \hspace{-0.1in} \eqref{eqn:LP1} - \eqref{eqn:LP4}, \slabel{eqn:LP_ED1}\\
	\rred{(p_{tk})} \hspace{-0.2in}	& &  C(1,k) \leq \sum_{s=1}^{k} \bigg( \lambda_s^+ \overline{C} - \lambda^-_s \underline{C} \bigg) + \overline{V} \mu_k  \nonumber\\
	&& + \sum_{s=2}^{k} V \bigg( \sigma_s^+ + \sigma_s^- \bigg) + \sum_{s=1}^{k}\sum_{j=1}^{N} \rered{b_j^s} \delta_{sj}, \nonumber \\
	&&   \forall k \in [t_0+1, T-1]_{\Z}, \slabel{eqn:LP_ED210}\\
	\rred{(p_{tk})} \hspace{-0.2in}	& &  C(t,k) \leq \sum_{s=t}^{k} \bigg( \lambda_s^+ \overline{C} - \lambda^-_s \underline{C} \bigg) + \overline{V} (\mu_t+\mu_k) \nonumber\\
	&& + \sum_{s=t+1}^{k} V \bigg( \sigma_s^+ + \sigma_s^- \bigg) + \sum_{s=t}^{k}\sum_{j=1}^{N} \rered{b_j^s} \delta_{sj}, \nonumber \\
	&&  \forall k \in [t+L-1, T-1]_{\Z}, \forall t \in [t_0+\ell+1, T-L]_{\Z}, \slabel{eqn:LP_ED2}\\
	\rred{(p_{tk})} \hspace{-0.2in}	&& C(t,T) \leq \sum_{s=t}^{T} \bigg( \lambda_s^+ \overline{C} - \lambda^-_s \underline{C} \bigg) + \overline{V} \mu_t \nonumber\\
	&& + \sum_{s=t+1}^{T} V \bigg( \sigma_s^+ + \sigma_s^- \bigg) + \sum_{s=t}^{T}\sum_{j=1}^{N} \rered{b_j^s} \delta_{sj}, \nonumber \\
	&&  \forall t \in [t_0+\ell+1, T]_{\Z}, \slabel{eqn:LP_ED22}\\
	\rred{(p_{tk})} \hspace{-0.2in}	& &  C(1,T) \leq \sum_{s=1}^{T} \bigg( \lambda_s^+ \overline{C} - \lambda^-_s \underline{C} \bigg) \nonumber\\
	&& + \sum_{s=2}^{T} V \bigg( \sigma_s^+ + \sigma_s^- \bigg) + \sum_{s=1}^{T}\sum_{j=1}^{N} \rered{b_j^s} \delta_{sj}, \slabel{eqn:LP_ED222}\\
	&& \eqref{eqn:EDd1} - \eqref{eqn:EDd5}, \nonumber\\
	&&   \forall t \in [t_0+\ell+1, T]_{\Z}, \forall k \in [\min \{t+L-1,T\}, T]_{\Z}.\slabel{eqn:LP_ED3}
\end{subeqnarray}

Note here that the {right-hand-sides} of constraints \eqref{eqn:LP_ED210} to \eqref{eqn:LP_ED222} correspond to the objective function \eqref{eqn:EDd_obj} under four different cases: 1) $t=1$, $k\in [t_0+1,T-1]_\Z$, 2) $t \in [t_0+\ell+1, T-L]_{\Z}$,$k \in [t+L-1, T-1]_{\Z}$, 3) $ t \in [t_0+\ell+1, T]_{\Z}$, $k = T$ and 4) the case in which $t=1$, $k = T$.

In the following, we \guan{present} an integral polytope for the original deterministic UC model \eqref{model:Nduc}. Before that, we take the dual of the above linear program \eqref{model:LP_ED} and obtain the following dual linear program:
\begin{subeqnarray} \label{model:LP_D1}
	&\hspace{-0.2in}\min  &\hspace{-0.1in}  \sum_{t=t_0}^{T-1} \textcolor{black}{S'}(t\yusub{+s_0})\textcolor{black}{w_t} + \hspace{-0.1in}\sum_{t=t_0+\ell+1}^{T-L}\sum_{k =t+L-1}^{T-1} \hspace{-0.1in} \textcolor{black}{S'}(k-t+1) \textcolor{black}{y_{tk}} \nonumber\\
	&& +\hspace{-0.1in} \sum_{kt \in \K\T} \hspace{-0.1in}\textcolor{black}{S}(t-k-1) \textcolor{black}{z_{kt}} +\hspace{-0.1in} \sum_{tk \in \T \K }  \sum_{s=t}^{k} \phi_{tk}^s  \nonumber \slabel{eqn:LP_D1obj}\\
	%&& \ \sum_{k=t_0}^{T-\ell-1}\sum_{t=k+\ell+1}^{T} \textcolor{black}{S} \textcolor{black}{z_{kt}} + \sum_{tk \in \T \K }  \sum_{s=t}^{k} \phi_{tk}^s  \slabel{eqn:LP_D1obj}\\
	&\hspace{-0.2in}\mbox{s.t.}  & \hspace{-0.1in}\sum_{t=t_0}^{T} \textcolor{black}{w_t} = 1, \slabel{eqn:LP_D11}\\
	&& \hspace{-0.1in}-\textcolor{black}{w_t} +\hspace{-0.15in} \sum_{k=t+\ell + 1}^{T}\hspace{-0.1in} \textcolor{black}{z_{tk}} -\hspace{-0.1in} \sum_{k=t_0+\ell+1}^{t-L+1}\textcolor{black}{y_{kt}} +\theta_t = 0, \forall t \in [t_0, T\hspace{-0.05in}-\hspace{-0.05in}1]_{\Z},\slabel{eqn:LP_D13}\\
	&& \hspace{-0.1in}	\sum_{k=\min\{t+L-1,T\}}^{T}\hspace{-0.1in}\textcolor{black}{y_{tk}} - \hspace{-0.1in}\sum_{k=t_0}^{t-\ell -1} \textcolor{black}{z_{kt}}= 0, \forall t \in [t_0+\ell+1, T]_{\Z},\slabel{eqn:LP_D12}\\
	&&\hspace{-0.1in} p_{tk} - \textcolor{black}{w_k} =0, \ \forall tk \in \T\K^1, \slabel{eqn:LP_D16}\\
	&&\hspace{-0.1in} p_{tk} - \textcolor{black}{y_{tk}} =0, \ \forall tk \in \T\K^2, \slabel{eqn:LP_D162}\\
	&& \hspace{-0.1in}q_{tk}^s \leq \overline{C}p_{tk}, \ \forall s \in [t,k]_{\Z}, \forall tk \in \T\K, \slabel{eqn:LP_D17}\\
	&& \hspace{-0.1in}-q_{tk}^s \leq - \underline{C}p_{tk}, \ \forall s \in [t,k]_{\Z}, \forall tk \in \T\K,   \slabel{eqn:LP_D18}\\
	&& \hspace{-0.1in}q_{tk}^t \leq \overline{V}p_{tk}, \ \forall tk \in \T\K^2, \slabel{eqn:LP_D19}\\
	&& \hspace{-0.1in}q_{tk}^k \leq \overline{V}p_{tk}, \ \forall tk \in \T\K, k\leq T-1 \slabel{eqn:LP_D191}\\
	&&\hspace{-0.1in} q_{tk}^{s-1} - q_{tk}^s \leq V p_{tk}, \  \forall s \in [t+1,k]_{\Z}, \forall tk \in \T\K, \slabel{eqn:LP_D110}\\
	&& \hspace{-0.1in}q_{tk}^{s} - q_{tk}^{s-1} \leq V p_{tk}, \  \forall s \in [t+1,k]_{\Z}, \forall tk \in \T\K, \slabel{eqn:LP_D111}\\
	&& \hspace{-0.1in}\phi_{tk}^s - a_j^s q_{tk}^{s} \geq \rered{b_j^s} p_{tk}, \hspace{-0.05in} \forall s \in [t,k]_{\Z}, \hspace{-0.01in}j \in  [1,N]_{\Z}, \forall tk \in \T\K, \slabel{eqn:LP_D112}\\
	&& \hspace{-0.1in}\textcolor{black}{w}, \textcolor{black}{z},p,y \geq 0, \theta_t \geq 0, \forall t \in [t_0,T-\ell-1]_\Z.
\end{subeqnarray}
%where $ \T\K $ represents the union of $\T\K ^1$ and $\T\K^2$. $\T\K^1$ represents the set of all possible combinations of $ t = 1$ and each $k \in [t_0+1,T]_\Z$ to construct a time interval $[t,k]_{\Z}$. $\T\K^2$ represents the set of all possible combinations of each $t \in [t_0+\ell+1, T-L]_{\Z}$ and each $k \in [t+L-1, T]_{\Z}$ to construct a time interval $[t,k]_{\Z}$. $ \K\T $ represents the set of all possible combinations of each $k \in [t_0, T-\ell -1]_{\Z}$ and each $t \in [k+\ell+1, T]_{\Z}$ to construct a time interval $[k,t]_{\Z}$.
%where $ \T\K $ represents the set of all possible combinations of each $t \in [1, T]_{\Z}$ and each $k \in [\min \{t+L-1,T\}, T]_{\Z}$ to construct a time interval $[t,k]_{\Z}$. 

In the above dual formulation, dual variables $ \textcolor{black}{w}$, $\textcolor{black}{z}$, $\theta $ and $y$ \rred{(labeled in the brackets of the \guan{left-hand-side} of \eqref{model:LP})} correspond to constraints \eqref{eqn:LP1} -- \eqref{eqn:LP4}, respectively, and dual variables $q$, $\phi$, and $p$ \rred{(labeled in the brackets of the \guan{left-hand-side} of \eqref{model:EDd} and \eqref{model:LP_ED})} correspond to constraints \eqref{eqn:LP_ED210} -- \eqref{eqn:LP_ED3} for each $ tk \in \T\K $, respectively.

%Note here that we can remove constraints \eqref{eqn:LP_D112} and {instead}, {recover} $\phi_{tk}^s$ {back to $ \phi_{tk}^s(q_{tk}^{s}, \textcolor{black}{y_{tk}}) $} as a general convex cost function of $q_{tk}^{s}$ and $\textcolor{black}{y_{tk}}$ {by letting the number of segments $ N $ for piecewise linear approximation go to infinity}. {More specifically, for the general convex cost case, the problem considered in \eqref{model:ED} can still be formulated as a convex program and thus strong duality holds.} In the following, {for modeling simplicity, }we remove constraints \eqref{eqn:LP_D112} and consider  $\phi_{tk}^s$ {to be} a general convex cost function as $ \phi_{tk}^s(q_{tk}^{s}, \textcolor{black}{y_{tk}}) $.

After replacing $ p $ with $ \textcolor{black}{w} $  or $\textcolor{black}{y}$ (due to \eqref{eqn:LP_D16} and \eqref{eqn:LP_D162}) in the dual formulation \eqref{model:LP_D1}, we obtain the cleaner model~\eqref{model:LP_D2} \guan{as described in Section~\ref{integralformulation}}.

Now, \gtwo{based on the above analysis,} we \guan{can} show that the polytope \eqref{eqn:LP_D21} --  \eqref{eqn:LP_D292} is an integral polytope in the following Theorem, which indicates that the extreme points of the polytope are integral. \rered{To prove this statement, we use the following \gtwo{proposition in~\cite{wolsey1998}}}.
	\begin{proposition}[\cite{wolsey1998}]
		For $A \in \R ^{m\times n}$ and $b\in \R ^m$, let 
		\begin{equation}
		\hspace{-0.1in}X=\{x \in \R^n: Ax \leq b, x_j \in \B, j \in J \subseteq \{1,\dots,n\}\}. \label{eq:proof}
		\end{equation}
		If $X$ is bounded, then the inequalities in~\eqref{eq:proof} describe the convex hull of $X$ if and only if for all $c\in \R^n$, the linear program $z^{LP} = \max\{cx: Ax \leq b\}$ has an optimal solution $x^* \in X$. 
	\end{proposition}

\begin{theorem} \label{integraltheorem}
	\guan{The alternative formulation~\eqref{model:LP_D2}, \gtwo{named} EUC, provides an integral formulation for the deterministic Unit Commitment problem~\eqref{model:Nduc}, i.e., leads to binary solutions \gtwo{with} respect to decision variables $w$, $y$, $z$ and $\theta$, by solving the problem as a linear program.}
\end{theorem} 

\begin{proof}
	\guan{Note here that, \gtwo{first} it is easy to observe that formulation~\eqref{model:LP_D2} is a valid formulation for the deterministic UC problem~\eqref{model:Nduc}. We now show that the extreme points for the polytope~\eqref{eqn:LP_D21}-\eqref{eqn:LP_D293} are binary with respect to decision variables $w$, $y$, $z$ and $\theta$.} 
	For notation brevity, we denote \eqref{eqn:LP_D2obj} as $\min c^{\top} (\textcolor{black}{w},\textcolor{black}{y},\textcolor{black}{z},\theta,q, \phi) $ where $ c $ is the column vector including all {coefficients} in \eqref{eqn:LP_D2obj}. Now we prove {that} for any value of $ c $, we can provide an optimal solution that is integral with respect to $\textcolor{black}{w}$, $\textcolor{black}{y}$, $\textcolor{black}{z}$, $\theta $ to the linear program with \eqref{eqn:LP_D2obj} as the objective function and \eqref{eqn:LP_D21}-\eqref{eqn:LP_D293} as constraints, which means that Theorem \ref{integraltheorem} holds.
	
	\gtwo{Following} the dynamic programming algorithm \eqref{new_dp1}-\eqref{new_dp2}, we can obtain an optimal solution for any given $c$. Based on this optimal solution, which indicates the online/offline status and generation amount of the {generator} at each time period, we construct a solution $(\textcolor{black}{w}^*,\textcolor{black}{y}^*,\textcolor{black}{z}^*,\theta^*,q^*, \phi^*)$ as follows:
	\begin{itemize}
		\item[1)] $ \textcolor{black}{w}^*_t =1$ if the {generator} shuts down for the first time at time $t+1$ and $\textcolor{black}{w}^*_t = 0$ otherwise, $\forall t \in [0,T-1]$.
		\item[2)] $ \textcolor{black}{w}^*_T = 1$ if the {generator} stays online for all the time periods and $\textcolor{black}{w}^*_T = 0$ otherwise.
		\item[3)] $ \textcolor{black}{y}^*_{tk} =1 $ if the {generator} starts up at time $ t $ and shuts down at time $ k + 1 $, and $ \textcolor{black}{y}^*_{tk} =0 $ otherwise.
		\item[4)] $ \textcolor{black}{y}_{tT}^* = 1$ if the {generator} starts up at time $t$ and stays online to the end and $\textcolor{black}{y}_{tT}^* = 0$ otherwise.
		%\item[3)] $ \varphi^*_t = 1$ if the {generator} shuts down at time $0$ and starts up at time $t$ and $\varphi^*_t = 0$ otherwise. 
		%\item[4)] $ \rho = 1$ if the {generator} shuts down for the first time at time $0$ and $\rho = 0$ otherwise. 
		\item[5)] $ \textcolor{black}{z}^*_{tk} =1 $ if the {generator} shuts down at time $ t + 1 $ and starts up at time $ k $ and $ \textcolor{black}{z}^*_{tk} =0 $ otherwise.
		\item[6)] $ \theta^*_{t} =1 $ if the {generator} shuts down at time $t + 1$ and stays offline to the end and $ \theta^*_{t} =0 $ otherwise.
		\item[7)] $q^{s*}_{tk}$ takes the value of optimal generation output for each $s \in [t,k]_{\Z}$ if the {generator} starts up at time $ t $ and shuts down at time $ k +1 $ and $q^{s*}_{tk} = 0$ otherwise, when $t > 1 $. It takes the value of optimal generation output for each $s \in [t,k]_{\Z}$ if the generator keeps on from the initial state and shuts down at time $k+1$ and $q^{s*}_{tk} = 0$ otherwise, when $ t = 1$.
		\item[8)] \textcolor{black}{$\phi^{s*}_{tk}$ takes the value of optimal \yusub{net} cost for each $s \in [t,k]_{\Z}$, which is \gtwo{equal to} $\max_{j\in [1,N]}\{a_j^sq_{tk}^{s*}+\rered{b_j^s}w_k\}$ when $tk \in \T\K^1$ and \gtwo{equal to} $\max_{j\in [1,N]}\{a_j^sq_{tk}^{s*}+\rered{b_j^s}y_{tk}\}$ when $tk \in \T\K^2$.}
		%\item[1)] $ \textcolor{black}{w}^*_{t} =1$ if the {generator} starts up for the first time at time $ t $ and $ \textcolor{black}{w}^*_{t} =0 $ otherwise.
		%\item[2)] $ \textcolor{black}{y}^*_{tk} =1 $ if the {generator} starts up at time $ t $ and shuts down at time $ k + 1 $ and $ \textcolor{black}{y}^*_{tk} =0 $ otherwise.
		%\item[3)] $ \textcolor{black}{z}^*_{tk} =1 $ if the {generator} shuts down at time $ t + 1 $ and starts up at time $ k $ and $ \textcolor{black}{z}^*_{tk} =0 $ otherwise.
		%\item[4)] $ \theta^*_{t} =1 $ if the {generator} shuts down at time $t + 1$ and stays offline to the end and $ \theta^*_{t} =0 $ otherwise.
		%\item[5)] $q^{s*}_{tk}$ takes the value of optimal generation output for each $s \in [t,k]_{\Z}$ if the {generator} starts up at time $ t $ and shuts down at time $ k +1 $ and $q^{s*}_{tk} = 0$ otherwise.
	\end{itemize}
	
	In the following, we show that $(\textcolor{black}{w}^*,\textcolor{black}{y}^*,\textcolor{black}{z}^*,\theta^*,q^*, \phi^*)$ is an optimal solution of model \eqref{model:LP_D2} with objective function \eqref{eqn:LP_D2obj} and constraints \eqref{eqn:LP_D21}-\eqref{eqn:LP_D293}.
	
	We first verify the feasibility. Since exact one $ \textcolor{black}{w}^*_t =1 $ for $t \in [t_0,T]_{\Z}$, constraint \eqref{eqn:LP_D21} is satisfied.
	For each $ t \in [t_0,T-1]_{\Z} $ in constraints \eqref{eqn:LP_D22}, we consider the following two possible cases: 
	\begin{enumerate}
		\item[1)] $\textcolor{black}{w}^*_t = 0$ and $ \textcolor{black}{y}^*_{kt} = 0 $ for all $k \in [t_0+\ell+1,T]_{\Z}$: \guan{for this case,} by definition, $ \theta^*_t=0 $ and $ \textcolor{black}{z}^*_{tk} = 0 $ for all possible $k \in [t+\ell+1,T]_{\Z}$.
		\item[2)] $\textcolor{black}{w}^*_t = 1$ or $\textcolor{black}{y}^*_{kt} = 1 $ for some $k \in [t_0+\ell+1,T]_{\Z}$: \guan{for this case}, the {generator} shuts down at time $ t+1 $. We can discuss this in two cases:
		\begin{enumerate}[label=(\roman*)]
			\item If $t \geq T-\ell$, $\sum_{k=t+\ell+1}^{T}\textcolor{black}{z}^*_{tk}+\theta^*_{t}$ reduces to $\theta_{t}$. In this case, the generator must satisfy the \guan{min}-down time constraints \gtwo{in the dynamic programming \yu{framework}, }and stay offline to the end, which indicates $\theta^*_t = 1$.
			\item If $t < T-\ell$, the generator either stays offline to the end, which indicates $ \theta^*_t =1 $ and $\textcolor{black}{z}^*_{tk} = 0$ for all possible $k \in [t+\ell+1,T]_{\Z}$, or starts up after satisfying the \guan{min}-down time limit, which indicates $\theta^*_t = 0$ and $\sum_{k=t+\ell+1}^{T} \textcolor{black}{z}^*_{tk} = 1$.
		\end{enumerate} 
	\end{enumerate}

	For both cases, constraints \eqref{eqn:LP_D22} are satisfied.
	
	For each $ t \in [t_0+\ell+1, T]_{\Z}$ in constraints \eqref{eqn:LP_D23}, we consider the following two possible cases:
	\begin{enumerate}
		\item[1)] $\textcolor{black}{z}^*_{kt} = 0$ for all $k \in [t_0, t-\ell-1]_{\Z}$: \gtwo{For this case, we have} $ \textcolor{black}{y_{tk}}^*  =0 $ for all $k \in [\min \{t+L-1,T\},T]_{\Z}$.
		\item[2)] $\textcolor{black}{z}^*_{kt} = 1$ for some $k \in [t_0, t-\ell-1]_{\Z}$: \gtwo{For this case, }the {generator} starts up at \gtwo{time} $t$. We can discuss this in two cases:
		\begin{enumerate}[label=(\roman*)]
			\item If $t > T-L$, $\sum_{k=\min \{t+L-1, T\}}^T \textcolor{black}{y}^*_{tk}$ reduces to $\textcolor{black}{y}^*_{tT}$. In this case, the generator must satisfy the \guan{min}-up time constraints and stay online to the end, which indicates $\textcolor{black}{y}^*_{tT} = 1$.
			\item If $t \leq T-L$, $\sum_{k=\min \{t+L-1, T\}}^T \textcolor{black}{y}^*_{tk}$ reduces to $\sum_{k=t+L-1}^{T}\textcolor{black}{y}^*_{tk}$. In this case, the generator either stays online to the end, which indicates $\textcolor{black}{y}^*_{tT} = 1$ and $\textcolor{black}{y}^{*}_{tk} = 0, \forall k \in [t+L-1,T-1]_{\Z}$, or shuts down again after satisfying the \guan{min}-up time limit, which indicates that $\sum_{k=t+L-1}^{T-1}\textcolor{black}{y_{tk}^*} =1$ and $\textcolor{black}{y}^*_{tT} = 0$.
		\end{enumerate}
	\end{enumerate}

	For both cases, constraints \eqref{eqn:LP_D23} are satisfied.
	
	\guant{Constraints \eqref{eqn:LP_D251}-\eqref{eqn:LP_D292}} are immediately satisfied \gtwo{based on} the construction of our solution and the definition of the economic dispatch problem. Also, \gtwo{constraints} \eqref{eqn:LP_D293} are satisfied obviously.
	
	We then verify the optimality. We claim that \gtwo{the value of} the objective function \eqref{eqn:LP_D2obj} \gtwo{for} the constructed solution equals to the objective value of the dynamic programming algorithm \eqref{new_dp1}-\eqref{new_dp2} {as follows}.
	\begin{subeqnarray}
		&& \hspace{-0.2in}\sum_{t=t_0}^{T-1} \textcolor{black}{S'}(t\yusub{+s_0})\textcolor{black}{w}^*_t + \hspace{-0.1in}\sum_{t=t_0+\ell+1}^{T-L}\sum_{k =t+L-1}^{T-1}\textcolor{black}{S'}(k-t+1) \textcolor{black}{y}^*_{tk} \nonumber\\
		&& +\hspace{-0.1in} \sum_{kt \in \K\T} \textcolor{black}{S}(t-k-1) \textcolor{black}{z}^*_{kt}  + \hspace{-0.1in}\sum_{tk \in \T \K }  \sum_{s=t}^{k}  \phi_{tk}^{s*} \nonumber\\
		%&&\sum_{t=1}^{T} SU(s_0 + t -1) \textcolor{black}{w_t}^* +  \sum_{t=1}^{T} \sum_{k=t+L-1}^{T-1} SD(k-t+1) \textcolor{black}{y_{tk}}^* + \sum_{t=L}^{T-\ell -1} \sum_{k=t+\ell+1}^{T} SU(k-t-1) \textcolor{black}{z_{tk}}^* \nonumber \\
		%&& \ + \sum_{t=T-\ell}^{T}E_t \theta_t^* + \sum_{tk \in \T \K }  \sum_{s=t}^{k} \bigg(a_{tk}^{s} q_{tk}^{s*} + b_{tk}^{s} \textcolor{black}{y_{tk}}^* \bigg)  \nonumber \\
		& \hspace{-0.2in}= & \hspace{-0.1in}\textcolor{black}{S'}(t_1\yusub{+s_0})  + \hspace{-0.5in} \sum_{tk \in \T\K ^2: \textcolor{black}{y_{tk}^*}=1, k\leq T-1, t \leq T-L} \hspace{-0.5in}\textcolor{black}{S'}(k-t+1) + \hspace{-0.3in} \sum_{kt \in \K\T: \textcolor{black}{z_{kt}^*}=1}\hspace{-0.2in} \textcolor{black}{S}(t-k-1) \nonumber\\
		&&   + \hspace{-0.2in}\sum_{tk\in \T\K^1: \textcolor{black}{w}^*_k = 1}\sum_{s=t}^{k}\phi_{tk}^{s*} + \sum_{tk \in \T\K^2 : \textcolor{black}{y_{tk}^*} =1 }  \sum_{s=t}^{k} \phi_{tk}^{s*} \slabel{pf1} \\
		& \hspace{-0.2in}= &\hspace{-0.1in} \textcolor{black}{S'}(t_1\yusub{+s_0})  + \hspace{-0.5in} \sum_{tk \in \T\K ^2: \textcolor{black}{y_{tk}^*}=1, k\leq T-1, t \leq T-L} \hspace{-0.5in}\textcolor{black}{S'}(k-t+1)  + \hspace{-0.3in} \sum_{kt \in \K\T: \textcolor{black}{z_{kt}^*}=1}\hspace{-0.2in} \textcolor{black}{S}(t-k-1)   \nonumber \\
		&&  + \sum_{tk\in \T\K^1: \textcolor{black}{w}^*_k = 1}C(t,k) + \sum_{tk \in \T\K^2: \textcolor{black}{y_{tk}^*} =1 }  C(t,k) \slabel{pf2}\\
		&\hspace{-0.2in} = &\hspace{-0.1in} V_\uparrow (0), \slabel{pf3}
	\end{subeqnarray}
	where $t_1$ in \eqref{pf1} \gtwo{indicates the first shut-down time, i.e.,} $ \textcolor{black}{w}_{t_1}^* = 1$. Equations \eqref{pf1} and \eqref{pf2} hold due to the construction of $\phi_{tk}^{s*}$. We discuss it in the following two cases:
	\begin{enumerate}
		\item \rered{When $y_{tk}^{*} = 0$ or $w_{k}^* = 0$, $q_{tk}^{s*} = 0$ due to constraints~\eqref{eqn:LP_D251}-\eqref{eqn:LP_D252}, which force $\phi_{tk}^{s*}=0$ in the minimization problem.} Based on the dynamic programming algorithm, we know $C(t,k)=0$ if the generator does not start up at time $t$ and shut down at time $k+1$. 
		\item When $y_{tk}^{*} = 1$ or $w_{k}^* = 1$, $\phi_{tk}^{s*}$ is forced to be $\max_{j\in [1,N]}\{a_j^sq_{tk}^{s*}+\rered{b_j^s}\}$ when $tk \in \T\K^1\cup \T\K^2$ in the minimization problem. Based on the dynamic programming algorithm, we know $C(t,k) = \sum_{s=t}^k \max_{j\in [1,N]}\{a_j^s x_s^* + b_j^s\}$. Since $x_s^* = q_{tk}^{s*}$ based on the construction, it is clear that $\sum_{tk\in \T\K^1: \textcolor{black}{w}^*_k = 1}C(t,k) =\sum_{tk\in \T\K^1: \textcolor{black}{w}^*_k = 1}\sum_{s=t}^{k}\phi_{tk}^{s*}, \sum_{tk \in \T\K^2: \textcolor{black}{y_{tk}^*} =1 }  C(t,k)=\sum_{tk \in \T\K^2 : \textcolor{black}{y_{tk}^*} =1 }  \sum_{s=t}^{k} \phi_{tk}^{s*} $. 
	\end{enumerate}   

	Equation \eqref{pf3} holds because $(\textcolor{black}{w}^*,\textcolor{black}{y}^*,\textcolor{black}{z}^*,\theta^*,q^*)$ is constructed based on the optimal solution of the dynamic programming algorithm \eqref{new_dp1} -- \eqref{new_dp2} {and is actually the expansion of the objective function in \eqref{new_dp2}}.
	By the Strong Duality Theorem, the constructed solution $(\textcolor{black}{w}^*,\textcolor{black}{y}^*,\textcolor{black}{z}^*,\theta^*,q^*,\phi^*) $ is an optimal solution for model \eqref{model:LP_D2}.
	
	From the above analysis, we notice that  $(\textcolor{black}{w}^*,\textcolor{black}{y}^*,\textcolor{black}{z}^*,\theta^*,q^*,\phi^*) $ is binary with respect to $ \textcolor{black}{w}, \textcolor{black}{y}, \textcolor{black}{z}$ and $\theta$ and optimal for the dual program for all possible cost coefficient $c$. 
	
	Thus, we have proved our claim.
\end{proof}
	
	\yu{Further, the construction of the solutions builds a bridge between the original space of the deterministic UC problem and the extended reformulation \eqref{model:LP_D2}. The following proposition can be derived directly.}
		\begin{proposition} \label{thm: opt solution_new}
			If  \gtwo{$(\textcolor{black}{w}^*,\textcolor{black}{y}^*,\textcolor{black}{z}^*,\theta^*,q^*,\phi^*) $} is an optimal solution to {the} dual program \eqref{model:LP_D2}, then
			\begin{eqnarray}
			&& \hspace{-0.3in}x_s^* = \sum_{tk \in \T\K, t\leq s \leq k} q^{s*}_{tk},u_s^* = \hspace{-0.1in} \sum_{tk\in \T\K^1, k \geq s} w^*_k+\hspace{-0.2in} \sum_{tk \in \T\K^2, t\leq s \leq k}\hspace{-0.2in} y^{*}_{tk},\nonumber\\
			&& \hspace{-0.3in} v_s^* = \sum_{kt \in \K\T, t= s} z^{*}_{kt}, \forall s \in [1,T]_{\Z} \slabel{eqn:opt solution_new} 
			\end{eqnarray}
			is an optimal solution to the deterministic UC problem \eqref{model:Nduc}.
		\end{proposition}

\begin{remark}
	If the generator is originally off for $s_0^-$ \guant{time units}, the earliest \guant{start-up} time will be $t_0^-$ with $t_0^{-} = [\ell-s_0^-+1]^+$, and we need to capture the first \guant{start-up} time $t$ instead of the first \guant{shut-down} time $t$ in the initial-on case. We \guant{can} redefine the binary decision variable $w_t$ \guant{to} represent whether the generator \guant{being offline for the whole operational horizon. We can apply the similar dynamic programming framework to obtain the corresponding formulation by the following adjustments}: %starts at time $t$ for the first time ($w_t = 1$) or not ($w_t = 0$) for $t=\{t_0^-,\dots, T\}$, and $w_{t_0^- - 1}$ represents the generator keeps offline for the while time periods. Meanwhile, we adjust the following:}
	\begin{enumerate}
		\item \yu{\guant{Change the} item considering $w_t$ in the objective function to $\sum_{t=t_0^-}^{T}S(s_0^-+t-1)w_t$.}
		\item \yu{\guant{Modify} the constraint~\eqref{eqn:LP_D21} as $\sum_{t=t_0^-}^{T}w_t = 1$}.
		\item \yu{\guant{Modify} the constraint~\eqref{eqn:LP_D22} as} 
		\begin{equation}
		\hspace{-0.1in}\sum_{k=t+\ell + 1}^{T}\hspace{-0.1in} \textcolor{black}{z_{tk}} -\hspace{-0.08in}\sum_{k=t_0^-}^{t-L+1}\hspace{-0.08in}\textcolor{black}{y_{kt}} +\theta_t = 0, \forall t \in [t_0^-\hspace{-0.05in}+L-1, T-\ell-1]_{\Z}
		\end{equation}
		\item \yu{\guant{Modify} the constraint~\eqref{eqn:LP_D23} as} 
		\begin{equation}
		\hspace{-0.1in}\sum_{k=\min\{t+L-1,T\}}^{T}\textcolor{black}{y_{tk}} -\hspace{-0.15in} \sum_{k=t_0^-+L-1}^{t-\ell -1} \hspace{-0.15in}\textcolor{black}{z_{kt}}-w_t = 0 , \forall t \in [t_0^-, T]_{\Z}
		\end{equation} 
		\item \yu{Since the \guant{initial} on-period sets $\T\K^1$ does not exist in initial off case, the constraints~\eqref{eqn:LP_D251}~\eqref{eqn:LP_D261}~\eqref{eqn:LP_D27}~\eqref{eqn:LP_D291} can be deleted, and $\T\K$ (as well as $\T\K^2$) represents all of the on-period time interval $[t,k]_\Z$, which contains all possible combinations of each pair of $t \in [t_0^-,T]_\Z$ and $k \in [\min\{t+L-1,T\},T]_\Z$.}
	\end{enumerate}
\end{remark}
\begin{remark}
\guant{Note here that our extended formulation approach can be applied to the case in which the problem parameters, including capacity and ramping rate, are dynamic.}~\yu{The symmetric ramping rates and constant parameters are assumed for symbolic simplification.}
\end{remark}

\section{Uplift Payments Minimization} \label{sec:uplift}
The system optimization problem for a $T$-period UC problem (with $\Lambda$ representing the set of generators) without considering transmission constraints run by an ISO can be abstracted as follows \yu{(defined as MEUC)}: 
\begin{eqnarray}\label{eq:1}
Z^*_{\tiny\mbox{QIP}}=& & \min_{\gred{w}^j, \gred{\psi}^j, \theta^j, \gred{y}^j, \gred{z}^j, \gred{q}^j} \sum_{j\in\Lambda}g_j(\gred{w}^j, \gred{\psi}^j, \theta^j, \gred{y}^j, \gred{z}^j, \gred{q}^j) \label{sys:model} \\
\textrm{s.t.} & & \sum_{j\in\Lambda} \sum_{tk\in\T\K} q^{j}_{tk}  =  d,  \label{loadbalance}  \\
&& (\gred{w}^j, \gred{\psi}^j, \theta^j, \gred{y}^j, \gred{z}^j, \gred{q}^j)  \in  \mathcal{X}^j_I, \ \forall  j\in\Lambda, \label{eq:f3}\\
&& \yu{w}^j, \gred{y}^j \ \mbox{and} \ \gred{z}^j \ \mbox{are binary}, \ \forall  j\in\Lambda, \label{con:inte}
\end{eqnarray}
where $\mathcal{X}^j_I$ is the feasible region for EUC formulation for generator $j$ with $\psi$ (mirror of $\phi$) being the generation cost. Let $Z^*_{\tiny\mbox{QP}}$ be the optimal objective value of the above formulation without the binary restriction constraints~\eqref{con:inte}. It is easy to observe that $Z^*_{\tiny\mbox{QIP}} \geq Z^*_{\tiny\mbox{QP}}$ because the latter is the objective value of a relaxed problem. 
%We denote its feasible region as $A$, where $A$ is a non-convex set, $z=(x_{1}, y_{1}, u_{1},..., x_{|\Lambda|}, y_{|\Lambda|}, u_{|\Lambda|})$and ISO's solution as $(x_j^*, y_j^*, u_j^*)$ for each generator.\\
Now we consider the profit maximization problem of each generator. For a given price vector $\pi$ offered by the ISO, the profit maximization problem for each generator can be described as follows: 
\begin{eqnarray}\label{eq:2}
\hspace{-0.4in} v_j(\gred{\pi})= & &\hspace{-0.4in} \mathop \max\limits_{\gred{w}^j, \gred{\psi}^j, \theta^j, \gred{y}^j, \gred{z}^j, \gred{q}^j} \hspace{-0.2in} {\gred{\pi}^T\sum_{tk\in\T\K} q^{j}_{tk}-g_j(\gred{w}^j, \gred{\psi}^j, \theta^j, \gred{y}^j, \gred{z}^j, \gred{q}^j)} \\
\textrm{s.t.}  & &  (\gred{w}^j, \gred{\psi}^j, \theta^j, \gred{y}^j, \gred{z}^j, \gred{q}^j)  \in  \mathcal{X}^j_I,\\
&& \yu{w}^j,\gred{y}^j \ \mbox{and} \ \gred{z}^j \ \mbox{are binary}. \label{cons:red3}
\end{eqnarray}
On the other hand, the profit generator $j$ can obtain following the ISO's schedule is equal to ${\gred{\pi}^T\sum_{tk\in\T\K} \bar{q}^{j}_{tk}-g_j(\gred{\bar{w}}^j, \gred{\bar{\psi}}^j, \bar{\theta}^j, \gred{\bar{y}}^j, \gred{\bar{z}}^j, \gred{\bar{q}}^j)}$, defined as $\bar{v}_j(\gred{\pi})$, where $(\gred{\bar{w}}^j, \gred{\bar{\psi}}^j, \bar{\theta}^j, \gred{\bar{y}}^j, \gred{\bar{z}}^j, \gred{\bar{q}}^j)$ is an optimal solution for generator $j$ in the system optimization problem corresponding to $Z^*_{\tiny\mbox{QIP}}$.

Since $v_j(\gred{\pi})$ is no smaller than $\bar{v}_j(\gred{\pi})$, there is a lost opportunity cost (LOC) of each generator following the ISO for each generator. \yu{Uplift payments} are triggered as described in~\cite{gribik2007market} and~\cite{hua2017convex} and can be represented as the following form:
\begin{eqnarray}\label{eq:3}
U_j(\gred{\pi}, \gred{\bar{w}}^j, \gred{\bar{\psi}}^j, \bar{\theta}^j, \gred{\bar{y}}^j, \gred{\bar{z}}^j, \gred{\bar{q}}^j)&=& \nonumber \\
& &\hspace{-1.8in} v_j(\gred{\pi}) -({\gred{\pi}^T\sum_{tk\in\T\K} \bar{q}^{j}_{tk}-g_j(\gred{\bar{w}}^j, \gred{\bar{\psi}}^j, \bar{\theta}^j, \gred{\bar{y}}^j, \gred{\bar{z}}^j, \gred{\bar{q}}^j)}).  \label{meq:15}
\end{eqnarray}
To reduce the discrepancy, we need to find an optimal price $\pi$ that minimizes the total uplift cost paid by the ISO. That is, we want to  
\begin{eqnarray}\label{eq:4}
 &\hspace{-0.1in}\min_{\pi} & \sum_{j\in\gred{\La}}U_j(\gred{\pi}, \gred{\bar{w}}^j, \gred{\bar{\psi}}^j, \bar{\theta}^j, \gred{\bar{y}}^j, \gred{\bar{z}}^j, \gred{\bar{q}}^j) \\
= &\hspace{-0.2in} \min_{\pi} &\hspace{-0.2in} \sum_{j\in\gred{\La}}(v_j(\gred{\pi})\hspace{-0.05in} -\hspace{-0.05in}({\gred{\pi}^T\hspace{-0.1in}\sum_{tk\in\T\K} \bar{q}^{j}_{tk}-g_j(\gred{\bar{w}}^j, \gred{\bar{\psi}}^j, \bar{\theta}^j, \gred{\bar{y}}^j, \gred{\bar{z}}^j, \gred{\bar{q}}^j)})) \label{eq:1} \\
=&\hspace{-0.2in} \min_{\pi} &\hspace{-0.2in} \sum_{j\in\gred{\La}} g_j(\gred{\bar{w}}^j, \gred{\bar{\psi}}^j, \bar{\theta}^j, \gred{\bar{y}}^j, \gred{\bar{z}}^j, \gred{\bar{q}}^j) - (\gred{\pi}^T d - \sum_{j\in\gred{\La}}v_j(\gred{\pi})),  \label{eq:2}
\end{eqnarray}
where~\eqref{eq:1} follows from~\eqref{meq:15} and~\eqref{eq:2} follows from~\eqref{loadbalance}. 
The above~\eqref{eq:2} is equivalent to solving the following maximization problem, since $\sum_{j\in\gred{\La}} g_j(\gred{\bar{w}}^j, \gred{\bar{\psi}}^j, \bar{\theta}^j, \gred{\bar{y}}^j, \gred{\bar{z}}^j, \gred{\bar{q}}^j)$ is a fixed value (the total generation cost for the system), as indicated in~\cite{gribik2007market}:
\begin{eqnarray}\label{eq:5}
 &\hspace{-0.1in}\max_{\pi} & \gred{\pi}^T d - \sum_{j\in\gred{\La}}v_j(\gred{\pi})  \nonumber \\
= &\hspace{-0.1in}\max_{\pi} & \gred{\pi}^T d -  \nonumber \\
&& \hspace{-0.5in}     \sum_{j\in\gred{\La}}(\mathop \max\limits_{\gred{w}^j, \gred{\psi}^j, \theta^j, \gred{y}^j, \gred{z}^j, \gred{q}^j} \hspace{-0.1in} {\gred{\pi}^T \hspace{-0.1in}\sum_{tk\in\T\K} q^{j}_{tk}}{-g_j(\gred{w}^j, \gred{\psi}^j, \theta^j, \gred{y}^j, \gred{z}^j, \gred{q}^j))}  \nonumber \\
& & \textrm{s.t.}  \hspace{0.2in} \ (\gred{w}^j, \gred{\psi}^j, \theta^j, \gred{y}^j, \gred{z}^j, \gred{q}^j)  \in  \mathcal{X}^j_I, \\
&& \hspace{0.4in} \yu{w}^j, \gred{y}^j \ \mbox{and} \ \gred{z}^j \ \mbox{are binary}. \label{cons:red2}\\
= &\hspace{-0.1in}\max_{\pi} & \sum_{j\in\gred{\La}}(\min_{\gred{w}^j, \gred{\psi}^j, \theta^j, \gred{y}^j, \gred{z}^j, \gred{q}^j} { g_j(\gred{w}^j, \gred{\psi}^j, \theta^j, \gred{y}^j, \gred{z}^j, \gred{q}^j)}  \nonumber \\
&& \hspace{0.5in} {- \gred{\pi}^T\sum_{tk\in\T\K} q^{j}_{tk} )} + \gred{\pi}^T d  \label{reobj} \\
& & \textrm{s.t.}  \hspace{0.2in} \ (\gred{w}^j, \gred{\psi}^j, \theta^j, \gred{y}^j, \gred{z}^j, \gred{q}^j)  \in  \mathcal{X}^j_I, \label{form:LR} \label{lrra}\\
&&\hspace{0.4in} \yu{w}^j, \gred{y}^j \ \mbox{and} \ \gred{z}^j \ \mbox{are binary}.  \label{cons:red}
\end{eqnarray}
It is easy to observe that model~\eqref{reobj}--\eqref{cons:red} is essentially the Lagrangian relaxation of the original problem \eqref{sys:model}-\eqref{con:inte} to obtain $Z^*_{\tiny\mbox{QIP}}$ and the corresponding optimal value $\pi$, i.e., $\pi^*$, is the optimal convex hull price.  

Based on Theorem~\ref{integraltheorem}, we can conclude that constraints~\eqref{cons:red3},~\eqref{cons:red2}, and~\eqref{cons:red} can be relaxed. Thus, since~\eqref{sys:model}-\eqref{eq:f3} is a linear program and~\eqref{reobj}-\eqref{lrra} is a Lagrangian relaxation of~\eqref{sys:model}-\eqref{eq:f3}, due to strong duality theorem, we can solve the linear program~\eqref{sys:model}-\eqref{eq:f3} and the optimal convex hull price $\pi^*$ is equal to the dual value corresponding to the load balance constraints~\eqref{loadbalance}. We highlight this main conclusion in the following theorem. 
\begin{theorem}
If the generation cost function $\bar f^s_{tk}(q_{tk}^{s})$ is convex and approximated by a piecewise linear function, then the optimal convex hull price can be obtained by solving the linear program \eqref{sys:model}-\eqref{eq:f3} and the optimal convex hull price is equal to the dual values corresponding to the load balance constraints~\eqref{loadbalance}.
\end{theorem}

%\section{An Example} 
\section{Case Study}\label{sec:exm}

%\begin{comment}
\yu{We first illustrate the MEUC formulation using an example in which two generators ($G1$ and $G2$) serve the loads in three periods ($d_1=40$MW, $d_2=80$MW, and $d_3=60$MW)}. For $G1$, there are no start-up cost and binary decisions. The generation bounds are $\Clower_1=0$ and $\Cupper_1=40$MW. The unit generation cost in each time period is $c_1=\$4/{\mbox{MWh}}, c_2=\$5/{\mbox{MWh}}$, and $c_3=\$6/{\mbox{MWh}}$. For $G2$, we have $\Clower_2=20$MW, $\Cupper_2=100$MW, $\overline{V}_2=55${\mbox{MW/h}}, $V_2=5${\mbox{MW/h}} and $L_2 = \ell_2 =2.$ The start-up cost for $G2$ is $\$100$. The convex generation cost for $G2$ is approximated by a two-piece piecewise linear function (e.g., $\psi \geq 20u + 4x$ and $\psi \geq -40u +5x$). We assume $s0 \geq L_2$ for $G2$, which indicates that the \guan{min}-up time constraints have been satisfied. 

We have \yu{the optimal objective value} $Z^*_{\tiny\mbox{QIP}}=\$835$ with the optimal solution $\bar{x}^1_1=0, \bar{x}^1_2=35,  \bar{x}^1_3=10, \bar{x}^2_1=40, \bar{x}^2_2=45,  \bar{x}^2_3=50$ for both the original 2-Bin and our \yu{MEUC} formulations. The corresponding LMPs are $\pi^1_1=1$, $\pi^1_2=5$, and $\pi^1_3=6$ (the optimal dual values corresponding to the load balance constraints in solving the economic dispatch problem when the unit commitment is fixed). Meanwhile, we have $Z^{\mbox{\tiny 2B}}_{\tiny\mbox{QP}}=\$808.18$ (the optimal objective value for the 2-Bin LP relaxation model) with the corresponding dual values $\pi^2_1=3.45$, $\pi^2_2=5$, and $\pi^2_3=5$ 
%(the corresponding fractional solution is $\hat{x}^1_1=40, \hat{x}^1_2=22.3,  \hat{x}^1_3=3.14, \hat{x}^2_1=30, \hat{x}^2_2=57.3,  \hat{x}^2_3=86.86$) 
and $Z^*_{\tiny\mbox{QP}}=\$828$ with the corresponding dual values $\pi^3_1=1.7$, $\pi^3_2=5$, and $\pi^3_3=6$ (the corresponding nonzero fractional solution is ${x}^1_1=0, {x}^1_2=36, {x}^1_3=12, q^{21}_{11}= 0, q^{21}_{12} =0, {q}^{21}_{13}=40, q^{22}_{12}= 0,{q}^{22}_{13}=44,  {q}^{23}_{13}=48, {q}^{23}_{33}=0, y^{2}_{33}=0, z^{2}_{03}=0, w^2_0=0.2, w^2_1=w^2_2=0, w^2_3 = 0.8, \theta^2_0 = 0.2, \theta^2_1=\theta^2_2 = 0, \theta^2_3 = 0.8$). Using $\pi^1$, $\pi^2$, and $\pi^3$ as the input for~\eqref{reobj}-\eqref{lrra}, combining~\eqref{eq:2}, we can obtain the uplift payments to be $\$35$, $\$26.82$, and $\$7$, respectively, which shows that the convex hull price provided by the \yu{MEUC} formulation leads to the smallest uplift payments and further, we only need to solve a linear program to achieve this.   
%\end{comment}

We further use a modified IEEE 118-bus system, based on the one given online at motor.ece.iit.edu/data, to \yu{to evaluate the performance of} the MEUC formulation. The system contains 54 generators and 118 buses. The formulation was coded in Python and solved using Gurobi 8.0.1. All experiments were implemented on a \yu{PC} with Intel Core i7-6500U CPU at 2.50GHz and 8GB memory. In our experiment, we let the operational horizon be 24 hours and the \gtwo{terminating mixed integer programming optimality gap be 0.01\%}. To generate more instances, corresponding to each nominal load \yu{$d_t, t=1,\dots,24$}, we construct 10 instances with the load for each instance uniformly distributed in $[0.9d_t, 1.1d_t]$. 

\yu{For each instance, we derive the price and the uplift payments under this price using three different pricing methods:
\begin{enumerate}
	\item TLMP method stated in Section~\ref{sec:intro}.
	\item Approximated CHP-Primal method proposed in~\cite{hua2017convex}.
	\item MEUC formulation~\eqref{sys:model}-\eqref{con:inte} defined in Section~\ref{sec:uplift}.
\end{enumerate}}

\yu{The computational results are reported in Table~\ref{tab:118case2}. In the table, Case 0 represents the nominal load case, and Cases $1-10$ represent the ten generated variations of the nominal load case. The uplift payments under TLMP method, approximated CHP-Primal method, and MEUC formulation are presented under the columns labelled ``$\text{U}_{\text{TLMP}}$(\$)", ``$\text{U}_{\text{ACHP}}$(\$)", and ``$\text{U}_{\text{MEUC}}(\$)$", respectively. The gaps between the uplift payments under MEUC formulation and that of the other two methods are calculated to show the improvement of MEUC formulation. Column ``$\text{Gap}_\text{TM}(\%)$" represents the gap between $\text{U}_{\text{TLMP}}$ and $\text{U}_{\text{MEUC}}$, and column ``$\text{Gap}_\text{CM}(\%)$" shows the gap between $\text{U}_{\text{ACHP}}$ and $\text{U}_{\text{MEUC}}$. More specifically, the gaps are calculated as follows:}
\begin{eqnarray*}
\text{Gap}_\text{TM} = \frac{\text{U}_{\text{TLMP}}-\text{U}_{\text{MEUC}}}{\text{U}_{\text{TLMP}}} \times 100\%\\
\text{Gap}_\text{CM} = \frac{\text{U}_{\text{ACHP}}-\text{U}_{\text{MEUC}}}{\text{U}_{\text{ACHP}}} \times 100\%
\end{eqnarray*}
%Table~\ref{tab:118case2} compares the uplift payment under the three pricing frameworks.  %The uplift payments under the EUC pricing scheme are equivalent to the difference between objective of UCED formulation and that of EUC formulation. For the LMP method and CHP-Primal formulation, the uplift payments are calculated as the difference of the maximized self-scheduling net revenues under the given price $\pi$ and the net revenue obtained by following the ISO's schedules. %Therefore, we first formulate and solve the self-profit maximization problem, based on the dual price $\pi$ obtained in TLMP or CHP-Prime problem, and label the optimal objective value as $P^*_{\tiny \text{self}}$. Then we calculate the net revenue following the ISO's schedule (labeled as $P^*_{\tiny\text{ISO}}$) as the difference between the profits ($\pi^Td$) and the cost, which is the optimal objective value of UCED problem $Z^*_{\tiny \text{QIP}}$. Thus, the uplift payments of TLMP or CHP-Prime formulation is calculated as $P^*_{\tiny \text{self}}-P^*_{\tiny\text{ISO}}$. We also present the savings (Saving(\$)) in the uplift payments by using EUC formulation comparied with the TLMP method.

% and we have proved that the EUC gap is the relative duality gap of the UCED problem and its Lagrangian dual in Section~\ref{sec:uplift}. 
%To show the tightness of EUC formulations, we first present the optimal objective values of UCED problem, CHP-Primal formulation, and EUC formulations in Table~\ref{tab:1181}.
\begin{table}[htbp]
	\centering
	\caption{The Computational Results for Modified 118 Cases}
	\begin{tabular}{cccccc}
		\toprule
		Case  & $\text{U}_{\text{TLMP}}$(\$)   & $\text{U}_{\text{ACHP}}$(\$) & $\text{U}_{\text{MEUC}}(\$)$ & $\text{Gap}_\text{TM}(\%)$ & $\text{Gap}_\text{CM}(\%)$ \\
		\midrule
			0 & 132  & 71  & 46  & 65.2   & 35.2   \\
			1 & 1480  & 775  & 259  & 82.5   & 66.6   \\
			2 & 1397  & 408  & 261  & 81.3   & 35.9   \\
			3 & 514  & 373  & 124  & 75.8   & 66.7   \\ 
			4 & 2714  & 608  & 244  & 91.0   & 59.9   \\
			5 & 4299  & 1478  & 616  & 85.7   & 58.3   \\
			6 & 2693  & 971  & 481  & 82.1   & 50.4   \\
			7 & 3293  & 1531  & 1143  & 65.3   & 25.3   \\
			8 & 952  & 356  & 90  & 90.5   & 74.6   \\
			9 & 4027  & 1064  & 549  & 86.4   & 48.4   \\
			10 & 5780  & 1038  & 671  & 88.4   & 35.3   \\
		\bottomrule
	\end{tabular}%
	\label{tab:118case2}%
\end{table}%
\yu{From Table~\ref{tab:118case2}, we can observe that for all cases, the MEUC formulation results in the least uplift payments. The large savings are observed between the TLMP method and the MEUC formulations because the TLMP method cannot incorporate the start-up and no-load costs of the generators. What is more, the uplift payments generated from the approximated CHP-Primal formulation are also larger than that from our proposed MEUC formulation, because the approximated CHP-Primal formulation only approximated the convex hull formulation for each UC since the ramping constraints and time-dependent start-up costs are not captured in their model. On the other hand, our MEUC formulation provides the convex hull description for each individual UC.} 

%\textcolor{green}{Note here that the uplift payments generated by our proposed MEUC model are small, which are also the minimal uplift payments, but they are not equal to zero, because non-zero uplift payments always exit in the non-convex UCED problem as stated in Section~\ref{sec:intro}. }	

\yu{Finally, we notice here that the uplift payments generated by our proposed MEUC model is small, but it is not equal to zero, because the integral formulation for the whole MEUC formulation is not obtained. We will perform future research along this direction.}

\bibliographystyle{IEEEtran}
%\bibliography{duc}
\bibliography{pricing}

\end{document}